\documentclass[final,5p,times,onecolumn]{elsarticle}
\usepackage{lineno, hyperref, amsmath, siunitx, color, graphicx, subcaption, caption, glossaries}
\journal{Nuclear Instruments \& Methods in Physics Research, Section A}
\glsdisablehyper            
\modulolinenumbers[1]       
\usepackage{booktabs}
\usepackage[T1]{fontenc}    
\usepackage{tablefootnote}

\usepackage{textcomp}       

\graphicspath{{Figures/}}

\begin{document}
    \begin{frontmatter}
        \title{Commissioning of low particle flux for proton beams at MedAustron}

        \author[hephy]{Felix Ulrich-Pur\corref{cor1}}
        \cortext[cor1]{Corresponding author}
        \ead{felix.ulrich-pur@oeaw.ac.at}
	    \author[MedAustron]{Laurids Adler}
        \author[hephy]{Thomas Bergauer} 
        \author[tu]{Alexander Burker}
        \author[MedAustron]{Andrea De Franco}
        \author[MedAustron]{Greta Guidoboni}
        \author[tu]{Albert Hirtl}
        \author[hephy]{Christian Irmler}
        \author[hephy]{Stefanie Kaser}
        \author[MedAustron]{Sebastian Nowak}
        \author[hephy]{Florian Pitters}
        \author[MedAustron]{Mauro Pivi}
        \author[MedAustron]{Dale Prokopovich}
        \author[MedAustron]{Claus Schmitzer}
        \author[MedAustron]{Alexander Wastl}

        \address[hephy]{Austrian Academy of Sciences, Institute of High Energy Physics (HEPHY),
        Nikolsdorfer Gasse 18, 1050 Wien, Austria}
		\address[tu]{TU Wien, Atominstitut, Stadionallee 2, 1020 Wien, Austria}
		\address[MedAustron]{MedAustron, Marie-Curie-Stra{\ss}e 5, 2700, Wiener Neustadt, Austria}
        \begin{abstract}
       		MedAustron is a synchrotron-based particle therapy centre located in Wiener Neustadt, Austria. It features three irradiation rooms for particle therapy, where proton beams with energies up to \SI{252.7}{MeV} and carbon ions of up to \SI{402.8}{MeV/u} are available for cancer treatment. In addition to the treatment rooms, MedAustron features a unique beamline exclusively for non-clinical research (NCR). This research beamline is also commissioned for proton energies up to \SI{800}{MeV}, while available carbon ion energies correspond to the ones available in the clinical treatment rooms. \\
Based on the requirements for particle therapy, all irradiation rooms offer particle rates of up to \SI{E9}{particles/s} for protons and \SI{E7}{particles/s} for carbon ions. However, for research purposes, lower particle fluxes are required and were therefore commissioned for the NCR beamline. Three particle flux settings with particle rates ranging from $\approx$ \SI{2.4e3}{particles/s} to $ \approx$ \SI{5.2e6}{particles/s} were established for seven proton energies below \SI{252.7}{MeV}. In addition to the particle rate, the spot sizes and beam energies were measured for these settings. Furthermore, three low flux settings for \SI{800}{MeV} protons with particle rates ranging from $\approx$ \SI{2e3}{particles/s} to $ \approx$ \SI{1.3e6}{particles/s} were commissioned. Since the commissioned low flux settings are in a regime well below the limits of the available standard beam diagnostics, setting up the beam under these new operational conditions entirely relied on the use of external detectors. Furthermore, a beam position measurement based alignment without using the standard beam profile monitors was performed for $\SI{800}{MeV}$ protons.
        \end{abstract}


    \end{frontmatter}


    \section{Introduction}
  	MedAustron is a particle therapy and research centre located in Wiener Neustadt, Austria. It features four irradiation rooms (IR1-IR4) with one beamline (IR1) exclusively dedicated to research \citep{Benedikt:67546}. Protons up to \SI{800}{MeV} and carbon ions of up to \SI{402.8}{MeV/u} can be delivered to this beamline. The design of the MedAustron accelerator complex was optimized by medical requirements \citep{Benedikt:1272158}, resulting in available particle rates in the order of $\approx \SI{E8}{}-\SI{E9}{particles/s}$ for medical use. However, experiments in non-clinical research often require measuring the interaction of single particles \citep{ULRICHPUR2020164407}, which leads to completely different beam and detector requirements compared to those for particle therapy.  Based on the accelerator design \citep{pimmms20002}, the maximum beam intensity could be theoretically lowered by a factor of 650, leading to a minimum particle rate of $\approx \SI{E6}{particles/s}$. But due to the rising demand for lower particle rates far below the design specifications \citep{pimmms20002} ($\approx \SI{E3}{}-\SI{E6}{particles/s}$), low flux settings for proton beams were commissioned in the research beamline at MedAustron. However, the beam for the required low flux settings had to be set up ``blindly'', as the conventional beam diagnostic elements were designed for use with therapeutic beam intensities down to $\approx \SI{E7}{particles/s}$.  Therefore, the lower limit of the available clinical intensities ($\approx \SI{E8}{particles/s}$), where beam instrumentation still could be used, was taken as the initial set-up for the low flux settings. From this point, the initial set-up was then extrapolated down to the low flux levels without the use of the standard beam instrumentation. Instead, to set up and monitor the low flux beam, external detectors were placed at the isocentre of IR1.

    \section{Materials and methods}
    \subsection{MedAustron accelerator}
\begin{figure*}[h]
\begin{center}
\includegraphics[width=\textwidth]{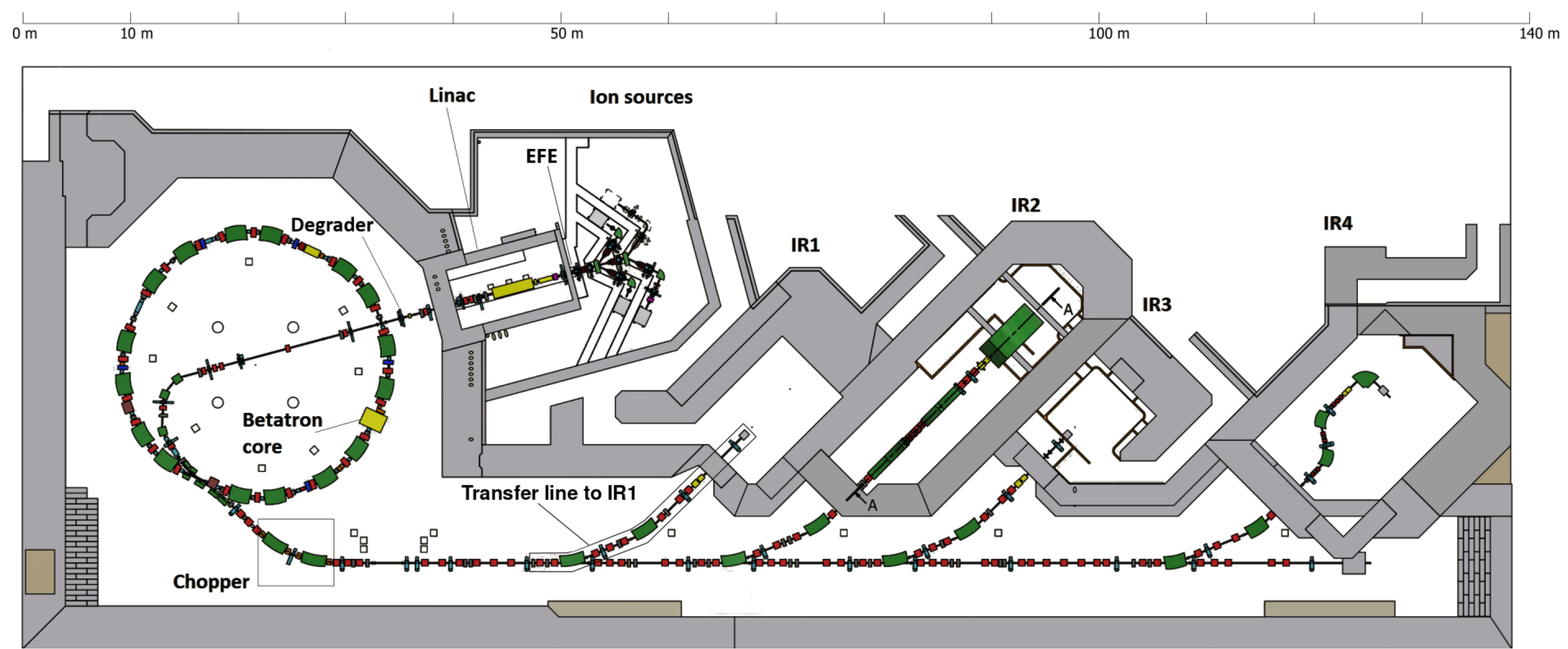}
\end{center}
\caption{Layout of the MedAustron accelerator complex where the key elements for low flux beam generation are shown. The (small) chopper magnets are located around the two HEBT dipoles in the box on the left. In the transfer line to IR1, the last seven quadrupoles are used to increase the transverse size of the beam for flux reduction.}
\label{fig:MAlayout}
\end{figure*}

    Figure \ref{fig:MAlayout} shows the MedAustron layout. As primary ions, ${\mathrm{H}_{3}}^+$ or ${^{12} \mathrm{C}^{4+}}$ are generated in two electron cyclotron resonance (ECR) ion sources at $\SI{8}{keV /u}$ \citep{Benedikt2011} located in the low energy beam transfer line (LEBT). The ions are then transported to the radio frequency quadrupole (RFQ) followed by an interdigital H-mode linear accelerator (LINAC), which accelerates the particles up to $\SI{7}{MeV \per u}$. After the LINAC, a carbon stripping foil is used to strip off remaining electrons and convert the ions to ${\mathrm{H}}^+$ or ${\mathrm{C}^{6+}}$. Those ions are then transported to the synchrotron by the medium energy beam transfer line (MEBT) and injected via a multi-turn injection scheme. After injection, the beam is bunched and accelerated by a radio frequency cavity. Inside the synchrotron ring ($\SI{77}{m}$ circumference), protons are accelerated up to $\SI{800}{MeV}$ and carbon ions up to $\SI{402.8}{MeV \per u}$.
Via a third order slow resonant extraction \citep{BENEDIKT1999523} the accelerated beam is then extracted towards one of the four irradiation rooms  through the High Energy Beam Transfer line (HEBT). 

  	\subsection{Particle flux reduction methods}
  	\label{sec:fluxreduction}
  	The flexible design of the MedAustron accelerator complex allows to expand the possible range of particle rates. Different methods were combined in order to reduce the particle rate per area. These are: 
  	\begin{enumerate}
  	    \item Reducing the number of particles injected into the synchrotron using the electrostatic fast deflector (EFE) or a degrader.
  	    \item Extending the extraction time into the HEBT to reduce the number of particles per second by changing the betatron core ramping.
  	    \item Scraping the beam into the chopper dump or increasing the transverse beam spot size in the HEBT so it exceeds the size of the vacuum tube. The increased size of the beam leads to a loss of particles in the HEBT and to a larger spot size at the isocentre, resulting in a reduced number of particles per unit area.
  	\end{enumerate}    It is important to notice that not all low flux settings mentioned above were applied for proton beams in the medical energy range (between 62.4 and 252.7 MeV) and at 800 MeV. The increased transverse beamsize method was not used in the case of 800 MeV to avoid environmental irradiation at that energy. In the following sections, a detailed description of the accelerator elements and the method used for flux reduction is given.
  	\label{sec:fluxmethods}
  	\subsubsection{Electrostatic fast deflector}
The EFE consists of two electrodes installed in the LEBT line. If a high voltage is applied (5.5 kV \citep{Borburgh:1407937}), the beam is deflected onto a cylindrical faraday cup, otherwise the beam can pass through towards the LINAC. The main purpose of the EFE is actually to adjust the pulse length: a typical injection scheme foresees $\SI{30}{}$-$\SI{50}{\mu s}$ pulses for the phase space painting in the synchrotron, and it can be triggered with a $\SI{1}{\mu s}$ resolution, therefore providing a well controllable measure to reduce the beam intensity arriving in the main ring.
    \subsubsection{Degrader}
  	In the MEBT section, a pepper pot like device (degrader) is used to adapt the number of particles injected into the ring \citep{degrader}. The beam intensity from the LINAC can be reduced to $\SI{10}{\%}$, $\SI{20}{\%}$ or $\SI{50}{\%}$ of the nominal value.
  	\subsubsection{Betatron core}
	The slow extraction mechanism used at MedAustron is the betatron core third order driven resonant extraction \citep{Pullia:IPAC2016-TUPMR037}. The use of the betatron core features the advantage of keeping all lattice elements constant during extraction and therefore the beam properties in the irradiation room.  
	
	At first, the beam is set slightly below the synchrotron's design energy as defined by the dipole's magnetic field. This causes the bunch to circulate along the dispersion orbit (the synchrotron RF and the resonant sextupole are located in a dispersion free region). During extraction, the machine is tuned such that the extraction occurs when the zero-betatron amplitude particle has reached the designed energy and thus circulates in the centre of the beam pipe.
	
	In preparation of the extraction, an RF jump is performed in order to create a more homogeneous $\Delta p/p$ distribution. Then the RF is turned off and a coasting beam is obtained after a few $\SI{}{ms}$. The resonant sextupole is ramped while the beam is still positioned safely away from the resonant tune. Finally, the beam is slowly driven into the resonance by inductively accelerating it with the betatron core. The acceleration effect is given by the flux change $\Delta \Phi = C \cdot B\rho \cdot \Delta p/p$ \citep{pimmms20001}, where $C$ is the machine circumference and $B\rho$ is the magnetic rigidity of the particle. As the beam is accelerated into the resonance, the large emittance particles become unstable first: their amplitude grows until they reach the electrostatic septum that deflects them into the extraction line. The betatron core ramping can be adjusted to extend the extraction time and reduce the particle rate. 
	
  	\subsubsection{Chopper}
\begin{figure}[h]
\begin{center}
\includegraphics[width=1.0\linewidth]{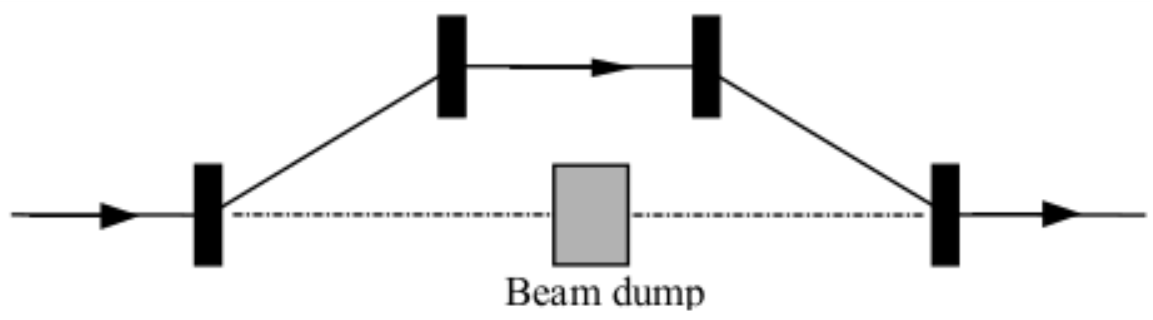}
\end{center}
\caption{Schematic of the chopper system. The black rectangles represent the four dipoles and the solid arrows show the beam path during normal operation. When the chopper is off, the beam hits the dump (grey box).}
\label{fig:Chopper}
\end{figure}
The chopper is a system located in the HEBT line (see Figure \ref{fig:MAlayout}) comprised of four dipoles powered in series that allow a fast (less than 300 ${\mu}$s) switching of the beam (controlled switch on/off of the beam during normal operations and for emergency interruptions) \citep{Benedikt2011}. Only when the dipoles are powered, the beam can circumvent the dump and be transmitted along the extraction line (see Figure \ref{fig:Chopper}). For low flux beam generation, the chopper has been used to control the transmission through the dump. The dipole strength was adjusted such that part of the beam was scraped on the dump and only a reduced portion of the extracted beam from the synchrotron reached the room.
This method has been applied to protons at 800 MeV only. 
	 	
  	\subsubsection{Quadrupole magnets to IR1}
\begin{figure}[h]
\begin{center}
\includegraphics[width=1.0\linewidth]{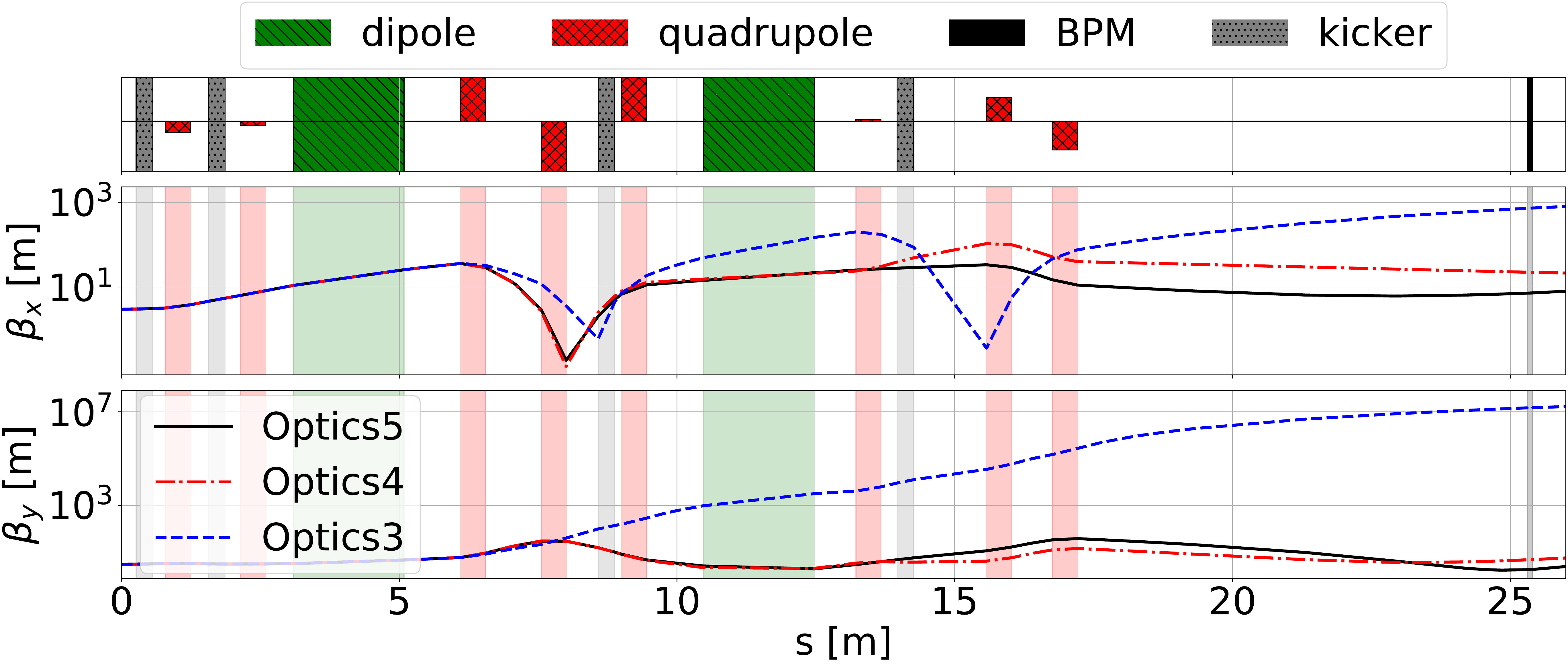}
\end{center}
\caption{On the upper side of the figure, details of the transfer line are shown. Quadrupole magnets are highlighted in red, dipole and kicker magnets in green and grey, respectively, and beam position monitors (BPM) in black. The $\beta$ functions along the transfer line to IR1 are displayed on the central (horizontal, $x$-plane) and bottom plots (vertical, $y$-plane) on a logarithmic scale. The nominal optics is shown in black, being very similar to Optics4 displayed in red. Optics3 is in blue and much larger in both planes compared to the other settings.}
\label{fig:betafunctions}
\end{figure}

  	The last 7 quadrupoles to IR1 - the irradiation room for non-clinical research purposes - were used to blow up the transverse (to the direction of motion) size of the beam and therefore reduce the particle flux at the monitor location. The modification of quadrupole settings implies a so-called optics change of the transfer line and the method will be referenced as optics adjustment in the following. Figure \ref{fig:betafunctions} shows a comparison of the $\beta_x$ and $\beta_y$ functions along the transfer line to IR1. The $\beta_x$ and $\beta_y$  functions represent the transverse beam envelopes which are related to the beam size in the transverse planes. For the same beam, a larger beta function implies a larger beam size. The quadrupole setting (Optics5) is used in case of the nominal setting in contrast to new settings with increased transverse beam size, Optics3 and Optics4, respectively.
  	
    \subsection{Low flux settings}
    \label{sec:fluxsettings}
    As mentioned in section \ref{sec:fluxreduction}, different methods for implementing low flux were applied depending on the beam energy. Table \ref{tab:settings} summarizes the combination of accelerator elements to achieve different particle rates ($\SI{}{particles \per \second}$), called ``Set. MED'' in case of medical energy range (between $\SI{62.4}{MeV}$ and $\SI{252.7}{}$ $\SI{}{MeV}$) or ``Set. p800'' for protons at $\SI{800}{MeV}$. For each energy range, three settings (``high'', ``medium'', ``low'') were commissioned. The order of magnitude of the obtained mean particle rate per setting defines the name of each setting and will be given in units of Hz. Setting ``high'', ``medium'' and ``low'' correspond to a mean particle rate of $ \mathcal{O} \left(\SI{}{MHz} \right)$, $ \mathcal{O} \left(\SI{100}{kHz} \right)$ and $ \mathcal{O} \left(\SI{}{kHz} \right)$, respectively.
\begin{center}
\begin{table}[h]   
\begin{tabular}{l c c c }
\toprule
 \textbf{Set. MED} & \textbf{high} & \textbf{medium} & \textbf{low} \\
 \midrule
 \textbf{Appr. rate} &$\SI{4.7}{MHz}$ & $\SI{350}{kHz}$ & $\SI{3}{kHz}$ \\
 \textbf{EFE} & $\SI{1}{\micro \second}$ & $\SI{1}{\micro \second}$ & $\SI{1}{\micro \second}$ \\
 \textbf{Degrader} & $\SI{10}{\percent}$ & $\SI{10}{\percent}$ & $\SI{10}{\percent}$ \\
  \textbf{Betatron} & Nom. & $\SI{10}{\percent}$ Nom. &$\SI{10}{\percent}$ Nom. \\
  \textbf{Optics} & Optics4 & Optics5 & Optics3 \\
  \textbf{H. size} $\mathbf{\sigma_x}$ & $\SI{3.9}{}$-$\SI{8.6}{mm}$& $\SI{3.2}{}$-$\SI{8.5}{mm}$ & $>\SI{5}{cm}$\\
    \textbf{V. size} $\mathbf{\sigma_y}$ & $\SI{2.8}{}$-$\SI{8.8}{mm}$& $\SI{2.4}{}$-$\SI{7.6}{mm}$ & $>\SI{5}{cm}$\\
  \toprule
 \textbf{Set. p800} & \textbf{high} & \textbf{medium} & \textbf{low} \\
 \midrule
 \textbf{Appr. rate} &$\SI{1.25}{MHz}$ & $\SI{200}{kHz}$ & $\SI{2}{kHz}$ \\
 \textbf{EFE} & $\SI{30}{\micro \second}$ & $\SI{10}{\micro \second}$ & $\SI{30}{\micro \second}$ \\
 \textbf{Degrader} & $\SI{20}{\percent}$ & $\SI{20}{\percent}$ & $\SI{20}{\percent}$ \\
  \textbf{Betatron} & Nom. &  Nom. &$\SI{12}{\percent}$ Nom. \\
  \textbf{Chopper} & $\SI{60.5}{\percent}$ Nom. & $\SI{60.5}{\percent}$ Nom. & $\SI{60.5}{\percent}$ Nom. \\
  \textbf{H. size} $\mathbf{\sigma_x}$\footnote[1]{}  & $\SI{8.2}{mm}$ & $\SI{9.4}{mm}$ & $\SI{8}{mm}$ \\
  


  \toprule
  \textbf{Set. NOM} & \textbf{nominal} & \\
  \midrule
 \textbf{Appr. rate} &$\SI{200}{MHz}$    & \\
 \textbf{EFE} & $\SI{30}{\micro \second}$ &  \\
 \textbf{Deggrader} & $\SI{10}{\percent}$    &\\
  \textbf{Betatron} & Nom.    &\\
  \textbf{Optics} & Optics5   &\\
  \textbf{H. size} $\mathbf{\sigma_x}$ & $\SI{2.8}{}$-$\SI{8.7}{mm}$& \\
   \textbf{V. size} $\mathbf{\sigma_y}$ & $\SI{2.9}{}$-$\SI{8.9}{mm}$& \\
  \bottomrule 
  \end{tabular}
   
  \begin{tabular}{l}
 \multicolumn{1}{l}{\hspace{-0.2cm}\footnote[1]{} Only the horizontal spot size $\sigma_x$ was measured for $\SI{800}{MeV}$} \\{(see section \ref{sec:spotsizemeas}).} 
\end{tabular}
 \vspace{-0.2cm}
\caption{The commissioned low flux settings are listed. For proton beams in the medical energy range, the settings include the combination of EFE, degrader and/or betatron core and optics adjustments. In case of protons at 800 MeV, the chopper is used instead of optics adjustment. The approximate mean rate and the horizontal and vertical spot size ($\sigma_x$, $\sigma_y$) are given for each setting. For comparison, the used nominal setting is also shown.}
\label{tab:settings}
\end{table}
\end{center}

	\subsection{Detectors used for beam characterization}
	The experimental set-ups for measuring the rate, spot sizes and beam energies are described in detail in the following section.
	\subsubsection{Rate monitor}
	Since the current beam diagnostics of MedAustron were not designed for low flux beams, a dedicated rate monitor (RaMon) to achieve single particle counting at low particle rates ($< \SI{50}{MHz}$) was built for the rate measurements. \\
For this purpose, two EJ228 plastic scintillators from Eljen \citep{eljen} were chosen as particle counters, since they have a rise and decay time of a few $\SI{}{ns}$. Each of the scintillators has a total volume of $\SI{50x50x10}{} \SI{}{{mm}^3}$. As light guide, PMMA fish tail lightguides ($\SI{50x10}{} \SI{}{{mm}^2}$) were connected to the photocathode ($\varnothing = \SI{8}{mm}$) of a Hamamatsu H10721-210 photosensor \citep{hamatsu}. Optical grease and double-sided optical duct tape were used for the assembly of each scintillator. The Hamamatsu photomultipliers (PMTs) were powered and readout by the AIDA 2020 trigger and logic unit (TLU) \citep{Baesso_2019}. A coincident signal of both PMTs defines a single particle count. Each coincident event was assigned with a timestamp using the internal $\SI{160}{MHz}$ clock of the TLU. Every $\SI{100}{ms}$, the number of coincidences and respective timestamps were processed by the Data Acquisition System (DAQ). For this purpose, the EUDAQ2 framework \citep{Liu_2019} was chosen since an interface for the AIDA TLU is already implemented in this software framework. An online beam monitor was developed using EUDAQ2 and QT \citep{QT} to calculate and display the particle rates. 30-50 spills were used per setting to calculate the average rate.
A schematic overview of the RaMon system can be seen in Figure \ref{fig:ds} on the right. 
	\subsubsection{Spot size measurement}
\begin{figure}
\begin{center}
\includegraphics[height=5cm]{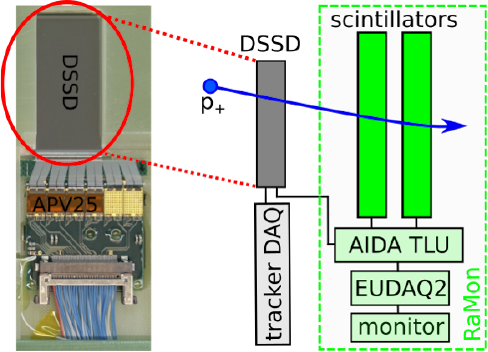}
\end{center}
\caption{Schematic overview of the spot size measurement for low flux protons using the DSSD and the RaMon system. On the left, an image of the p-side ($y$-coordinate) of the DSSD is shown. The DSSD strips are connected to 4 APV25 chips mounted on hybrid boards which are readout with a custom DAQ system. The RaMon plastic scintillators are connected to the AIDA2020 TLU and are used to trigger single particle events.}
		
		
		\label{fig:ds}
\end{figure}

\begin{figure}[h]
	\begin{center}
	\includegraphics[width=0.25\textwidth]{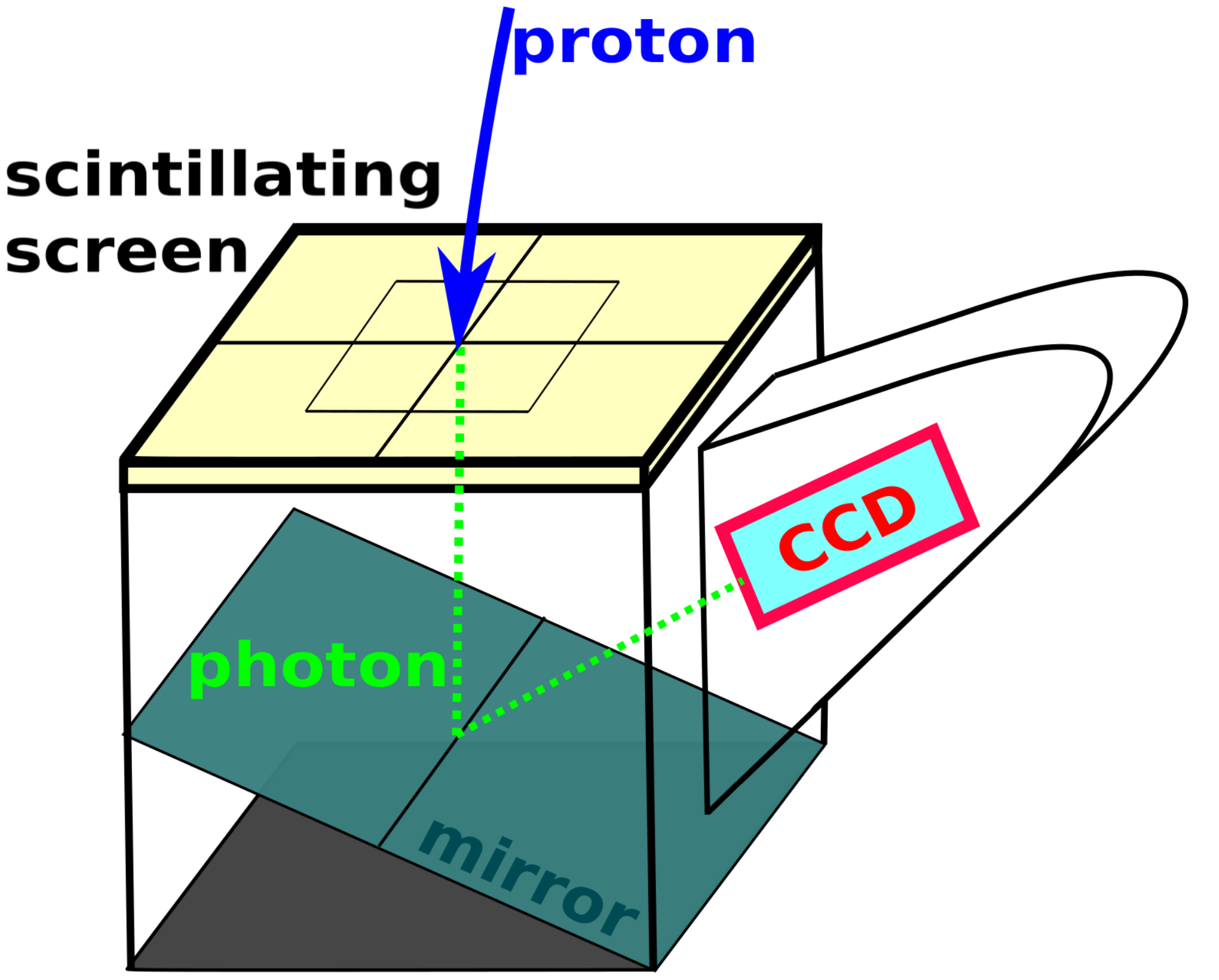}
	\end{center}
	\caption{Schematic overview of the spot size measurement with the Lynx\textsuperscript{\textregistered} detector. The interaction position of an incident particle on the $\SI{30}{}\times \SI{30}{{cm}^2}$ scintillating screen is mapped onto the $\SI{600}{} \times \SI{600}{}$ pixel of a CCD camera via a mirror.}
	\label{fig:lynxsetup}
\end{figure}

	Depending on the particle flux, two different detectors for measuring the spot size at the IR1 isocentre were used. \\
 For all low flux settings, a double sided silicon strip detector (DSSD), which has already been used for single particle tracking at MedAustron \citep{ULRICHPUR2020164407}, was used. The n-substrate based DSSD is $\SI{300}{\micro m}$ thick and has an active area of $\SI{2.56}{} \times \SI{5.12}{{\centi \metre}^2}$. Each side features $\SI{512}{}$ orthogonal AC coupled strips with a pitch of $\SI{100}{\micro \metre}$ on the n-side ($x$-coordinate) and $\SI{50}{\micro \metre}$ on the p-side ($y$-coordinate). The sensor itself is readout via 4 APV25 chips \citep{APV25} on each side ($\SI{128}{\mathrm{strips/APV}}$, Figure \ref{fig:ds}). A custom readout system \citep{ULRICHPUR2020164407}, which was originally developed for the Belle II Silicon Vertex Detector \citep{BelleIISVDReadout}, was used. The readout was triggered using the RaMon plastic scintillators, which were placed behind the DSSD.  A schematic overview of the experimental setup is shown in Figure \ref{fig:ds}.	
\\Due to the limited particle rate capability of the RaMon system ($< \SI{50}{MHz}$), the Lynx\textsuperscript{\textregistered} detector (IBA-Dosimetry, Schwarzenbruck) was used for the nominal setting ($\approx\SI{200}{MHz}$). This detector has already been used for spot size and position verification at clinical beam intensities during commissioning in the other treatment rooms \citep{STOCK2018196}, \citep{Grevillot2019}. The Lynx\textsuperscript{\textregistered} detector itself is a gadolinium-based plastic scintillating screen with an active area of $\SI{30}{}\times \SI{30}{{cm}^2}$ and a thickness of $\SI{0.4}{mm}$. Via a mirror, the scintillating screen is coupled to a CCD camera with $\SI{600}{} \times \SI{600}{}$ pixel and a pixel size of $\SI{4.25}{} \times \SI{4.25}{\mu m^2}$\citep{RUSSO201748}. All components of the Lynx\textsuperscript{\textregistered} detector are contained in a light-tight box  with $\SI{360}{}\times \SI{370}{}\times \SI{600}{\milli m^3}$. A schematic overview of the Lynx\textsuperscript{\textregistered} is depicted in Figure \ref{fig:lynxsetup}. 
\mbox{}\\
Whenever a particle deposits energy in the scintillating screen, the deposited energy is converted into green light ($\SI{540}{nm}$), which is then reflected towards a CCD camera  via a mirror. The amount of light reaching the CCD camera can be modified by an aperture collimator system (iris) to avoid saturation. Furthermore, the exposure time for a single image can be adapted ($ \leq \SI{180}{s}$). According to \citep{RUSSO201748}, the effective spatial resolution of the Lynx\textsuperscript{\textregistered} detector is $\SI{0.5}{mm}$. Even though the Lynx\textsuperscript{\textregistered} detector could also have been used for particle rates down to $\mathcal{O} \left( \SI{100}{kHz}\right)$, the DSSD system was used instead for all low flux settings due to its superior spatial resolution.
\subsubsection{Energy measurement}	
\begin{figure}[!h]
	\begin{center}
	\includegraphics[width=0.4\textwidth]{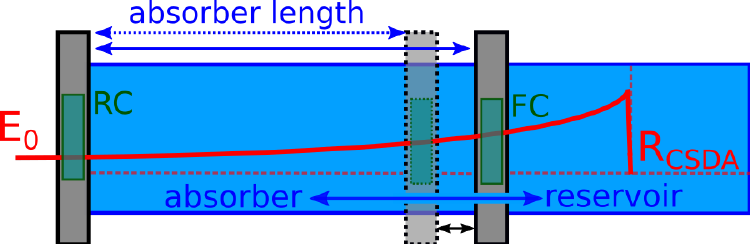}	
	\end{center}
	\caption{The PTW PEAKFINDER\texttrademark{} consists of two interconnected, water-filled bellows (absorber and reservoir) and 2 ionization chambers (RC, FC). The depth-dose profile in water was obtained by measuring the dose to water at different positions of the FC in the water column. From the sampled depth-dose profiles, the range in water was derived for different flux settings and beam energies.}
	\label{fig:peakfinder}
	\end{figure}

In addition to the spot size and rate, the influence of each low flux setting on the beam energy was studied by measuring the range in water for different beam energies and flux settings. \\The range $R$ in a particular material can be obtained using the continuous-slowing-down approximation (CSDA). In order to calculate the CSDA range for a proton with initial energy $E_0$, the reciprocal total stopping power $S(E)$ has to be integrated between the proton's initial state $E\!\!=\!\!E_0$ and stationary state $E\!\!=\!\!0$.
\begin{linenomath}
\begin{equation}
	R_{\mathrm{CSDA}}=\int^0_{E_0} {\left(-\frac{\mathrm{d}E}{\mathrm{dx}} \right) }^{-1} \mathrm{d}E=\int^0_{E_0} \frac{\mathrm{d}E}{S(E)}
	\label{eq:csda}
	\end{equation}
\end{linenomath}
For a homogeneous material, simplified, parameterized versions of equation (\ref{eq:csda}) (\citep{braggkleemann}, \citep{Donahue_2016}) can be used to convert the measured range $R_{\mathrm{CSDA}}$ into the initial beam energy $E_0$. Therefore, the range obtained in a residual range detector is usually expressed in water equivalent thickness (WET\citep{Zhang2009}). If the traversed material in the range detector consists of $n$ different, homogeneous materials, with total stopping powers $S_i (E)$ ($i \in [1,n]$) and the traversed pathlength per material is $\mathrm{t}_i$, equation (\ref{eq:csda}) can be rewritten as follows
\begin{linenomath}
	\begin{flalign}
	R_{\mathrm{CSDA}} &= \underbrace{\int^{E_1}_{E_0} \frac{\mathrm{d}E}{S_1(E)}}_{\hat{=} \mathrm{t_1}}+\underbrace{\int^{E_2}_{E_1} \frac{\mathrm{d}E}{S_2(E)}}_{\hat{=} \mathrm{t_2}}+ \ ..+\underbrace{\int^{0}_{E_\mathrm{n-1}} \frac{\mathrm{d}E}{S_n(E)}}_{\hat{=} \mathrm{t}_n} 
	\\ \noindent \Rightarrow R_{\mathrm{CSDA}} &= \!\sum_i^n \mathrm{t}_i.
	\end{flalign}
\end{linenomath}
In order to convert the measured CSDA range into range in water, the water equivalent thickness $\mathrm{WET}_i$ of each traversed pathlength $\mathrm{t}_i$ leading to $R_{\mathrm{CSDA}}$ has to be calculated according to \citep{Zhang2009} 
\begin{linenomath}
\begin{equation}
\mathrm{WET}_i \approx \mathrm{t}_i \frac{S_{\mathrm{i}}(E)}{S_{\mathrm{H_2O}}(E)} \Rightarrow R_{\mathrm{CSDA,WET}} = \sum_i^n \mathrm{WET}_i.
\label{eq:wet}
\end{equation}
\end{linenomath}
For the following measurements, the total stopping power in water $S_{\mathrm{H_2O}}(E)$, as well as the total stopping power for each traversed material $S_{\mathrm{i}}(E)$ were taken from the NIST database \citep{berger2017stopping} to obtain the final residual range in water $R_{\mathrm{CSDA,WET}}$.  \\
For particle rates above $\mathcal{O} \left(\SI{100}{kHz} \right)$, the residual range in water was measured directly using the PTW PEAKFINDER\texttrademark{} (PTW, Freiburg, Germany). The PTW PEAKFINDER\texttrademark{} (Figure \ref{fig:peakfinder}) is a height-adjustable water column with a diameter of $\SI{114}{mm}$ and two ionization chambers to measure the depth dose profile of a particle beam in water. The water column itself consists of two interconnected water-filled bellows (absorber and reservoir). The PTW TM34082 is used as a reference chamber (RC) in front of the absorber water column and the PTW TM34080 as a field chamber (FC) between the absorber and reference water column. The position of the FC as well as the length of each bellow is adjusted via a servo motor (stepsize $\geq\!\SI{10}{\mu m}$). To sample the depth-dose profile, the absorbed dose to water \citep{international2001iaea} in the FC is measured relative to the absorbed dose to water in the RC at different positions inside the water column ($\SI{340}{mm}$ movable range). 

Close to the maximum (Bragg peak), the depth-dose profile was sampled in $\SI{0.3}{mm}$ steps. The range in water was then measured at the $\SI{20}{\%}$ distal fall-off relative to the Bragg peak for each energy and compared to the ranges from the NIST PSTAR database. 
	\\
	Due to a low signal-to-noise ratio in the ionization chambers below particle rates of $\mathcal{O} \left(\SI{100}{kHz} \right)$, no depth-dose profiles could be obtained with the PTW PEAKFINDER\texttrademark{}. Instead, a range telescope, formerly developed by the TERA foundation \citep{Bucciantonio}, was used to measure the particle range in water. Fast silicon photomultipliers (Hamamatsu MPPC S10362-11-050C \citep{sipm}) coupled to 38 plastic scintillator slices with a water equivalent thickness of $\SI{3.6}{mm}$ and an active area of $\SI{30}{} \times \SI{30}{{\centi \metre}^2}$ each allow range measurements of single particles. The readout of a single particle event is triggered by the coincident signal of the RaMon plastic scintillators placed in front of the range telescope. The measured range in the telescope is then converted to residual range in water according to equation (\ref{eq:wet}). As described in \citep{ULRICHPUR2020164407} and \citep{dissbuccantonio}, the range telescope suffered from severe voltage instabilities, therefore the mainboard and the readout software were completely redesigned and replaced prior to the measurements.
\begin{figure}[h]
\begin{center}
\includegraphics[width=0.4\textwidth]{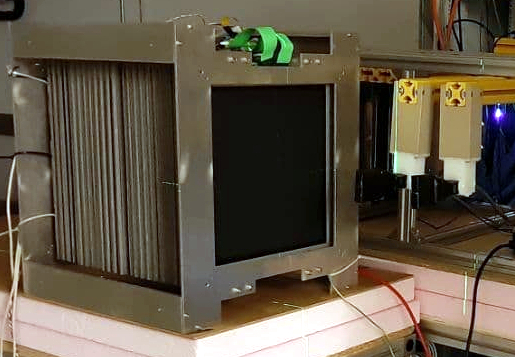}
\end{center}
\caption{Range measurement for low flux protons at MedAustron using the TERA range telescope with two plastic scintillators in front as triggers.}
\label{fig:tera}
\end{figure}
\subsection{Additional challenges}
\label{sec:steering}
For normal (medical) operational conditions, the beam position and profile in the HEBT can be measured with scintillating fibre hodoscopes (SFX) \citep{Feurstein:2012zz}. Due to the low beam current of the low flux proton beams, the SFX detectors in the extraction line do not produce enough signal to measure a beam profile, effectively rendering the transfer line ``blind'' to the beam position until the isocentre.  Therefore, under these new operational conditions, the verification of the beam position and beam angle at the isocentre completely relied on external detectors. In order to measure the beam angle at the isocentre, three additional DSSDs were placed behind the first DSSD at the isocentre at equal distances ($\SI{10}{cm}$). The measured angles were found to be negligible for all settings.
\subsubsection{Steering of the $\SI{800}{MeV}$ low flux beam}
During commissioning of the $\SI{800}{MeV}$ low flux beam, apart from a reduced strength of the chopper dipoles, the setpoints of all other elements in the extraction line were initially left unchanged compared to the nominal flux beam. During measurement of the spot sizes with the DSSD, an offset of the beam in the isocentre of $\SI{+12.5}{mm}$ horizontally and $\SI{-13}{mm}$ vertically could be observed. In order to correct the beam position at the isocentre while not changing the beam angle, a second position measurement upstream would be required. Since that proved to be impossible due to the low beam current, an adapted steering strategy was employed, where the corrector kick response matrix of the beam position \citep{Wiedemann2015}, using the last two horizontal and vertical correctors upstream the IR1 isocenter, was measured for the nominal flux beam on the SFX close to the last quadrupole of the beamline and at the isocentre. The kick response matrix describes the change in beam position as a function of a difference in applied kick angle $\Delta {\Theta}_{x,j}$ by a corrector magnet. The elements of the kick response matrix are defined as (exemplary for the horizontal plane)\citep{Wiedemann2015}

\begin{equation}
 C_{ij}^{xx}=\frac{\Delta x_i}{\Delta {\Theta}_{x,j}} \left[\SI{}{m \per rad} \right],
 \end{equation}
 
 where $\Delta x_i$ is the change of the horizontal beam position at the beam profile monitor $\mathrm{BPM}_i \left[\SI{}{m} \right]$ and $\Delta {\Theta}_{x,j}$ is the change of the horizontal kick angle of corrector $j \left[\SI{}{rad} \right]$. After obtaining the kick response matrix via measurement of the position response, the required change in corrector kick angles $\Delta \vec{\Theta}_{x_{needed}}$ to correct the beam position on all beam profile monitors to its target position (reference trajectory) can be calculated by inverting the kick response matrix
 
 \begin{gather}
 \begin{pmatrix} \Delta \vec{\Theta}_{x_{\mathrm{needed}}}  \\ \Delta \vec{\Theta}_{y_{\mathrm{needed}}} \end{pmatrix}
 ={ 
  \begin{pmatrix}
   C^{xx} & C^{xy} \\
   C^{yx} & C^{yy}  
   \end{pmatrix}}^{-1} \cdotp \ \begin{pmatrix} \Delta \vec{x}_{\mathrm{target}}  \\ \Delta \vec{y}_{\mathrm{target}} \end{pmatrix}.
\end{gather}
Where $\Delta \vec{x}_{\mathrm{target}}$ is the vector of the difference between the (horizontal) beam position and the target beam position on all beam profile monitors. The inverted kick response matrix of the nominal flux beam was then applied to the measured offset of the low flux beam, resulting in new corrector strengths compensating the offset while still minimizing the beam angle at the isocentre. Without being able to measure the low flux beam with the $\mathrm{BPM}_i$ in the HEBT, this beam position measurement based alignment was performed using the DSSDs at the isocentre only. With this method, the $\SI{800}{MeV}$ low flux beam was aligned prior to the rate and spot size measurements with $\SI{800}{}$ $\SI{}{MeV}$ protons.
    \section{Results}
    As described in Section \ref{sec:fluxsettings}, three different flux settings could be commissioned for beam energies below $\SI{252.7}{MeV}$ and three different settings for $\SI{800}{MeV}$. For details of the settings, the reader is referred to Table \ref{tab:settings}.

    \subsection{Rate measurement}
\begin{figure}[h]
	\centering
	\includegraphics[width=0.5\textwidth]{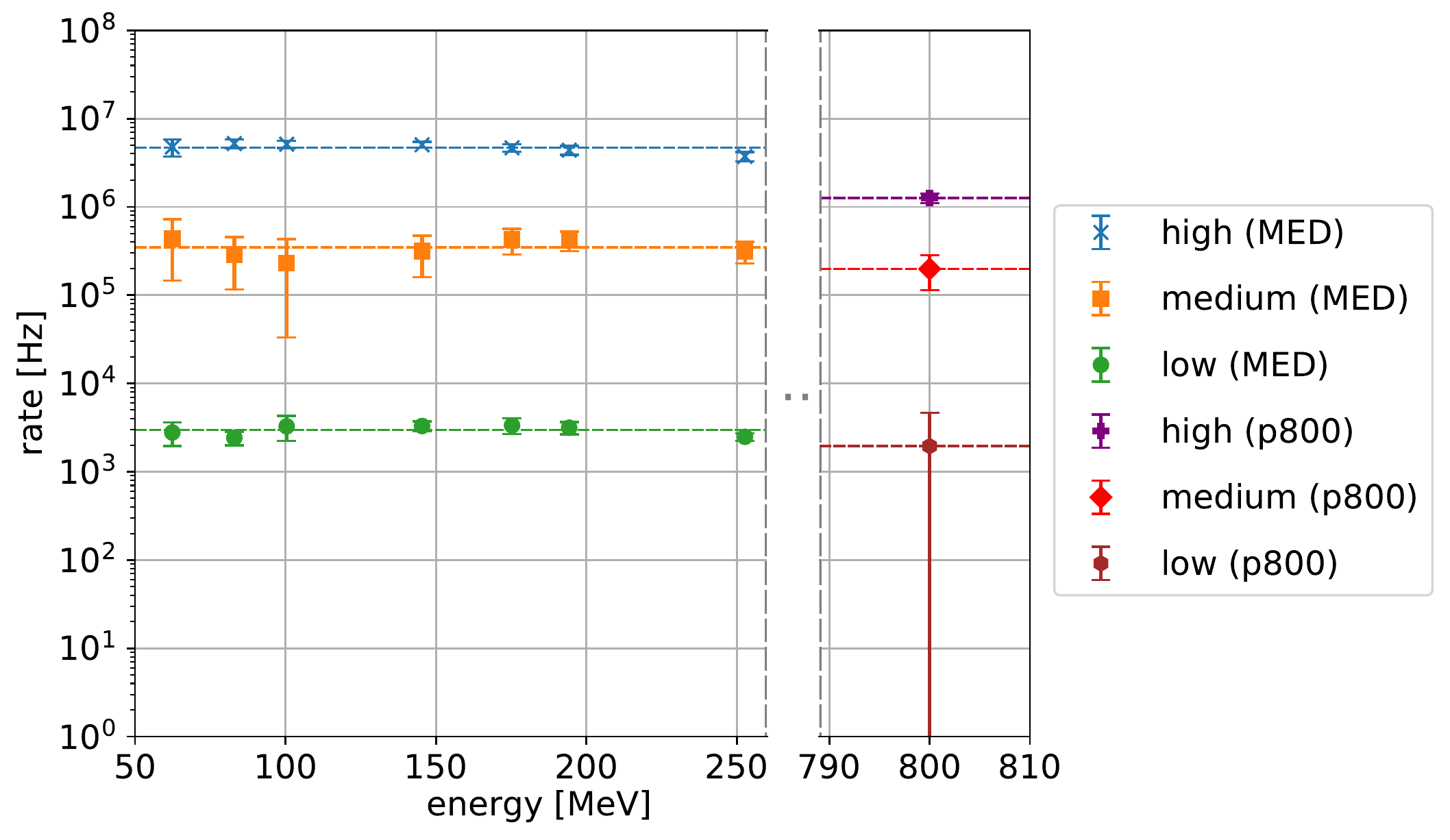}
	\caption{Resulting low flux particle rates measured with the RaMon system. Three different flux settings per energy could be commissioned. The settings for $\SI{800}{MeV}$ protons differed from the medical settings. The dashed horizontal lines represent the mean particle rate per setting, which also defines the name of the settings.}
	\label{fig:rateresults}
	\end{figure}

\begin{figure}[!h]
		\centering
		\begin{subfigure}{.5\linewidth}
		\includegraphics[width=1.\linewidth]{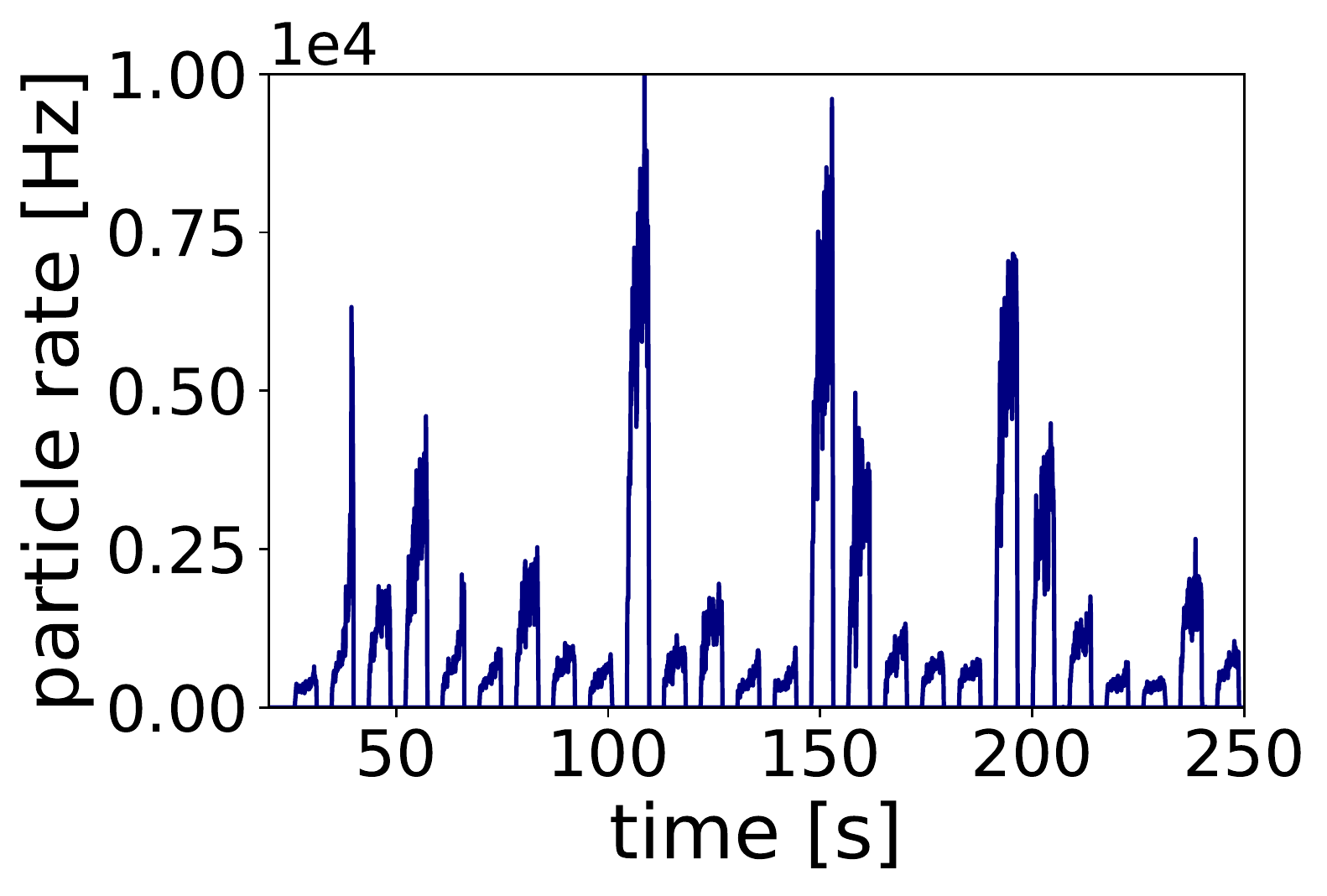}
		\caption{``low (p800)'' setting}		
		\end{subfigure}%
		\begin{subfigure}{.5\linewidth}
		\centering
		\includegraphics[width=0.92\linewidth]{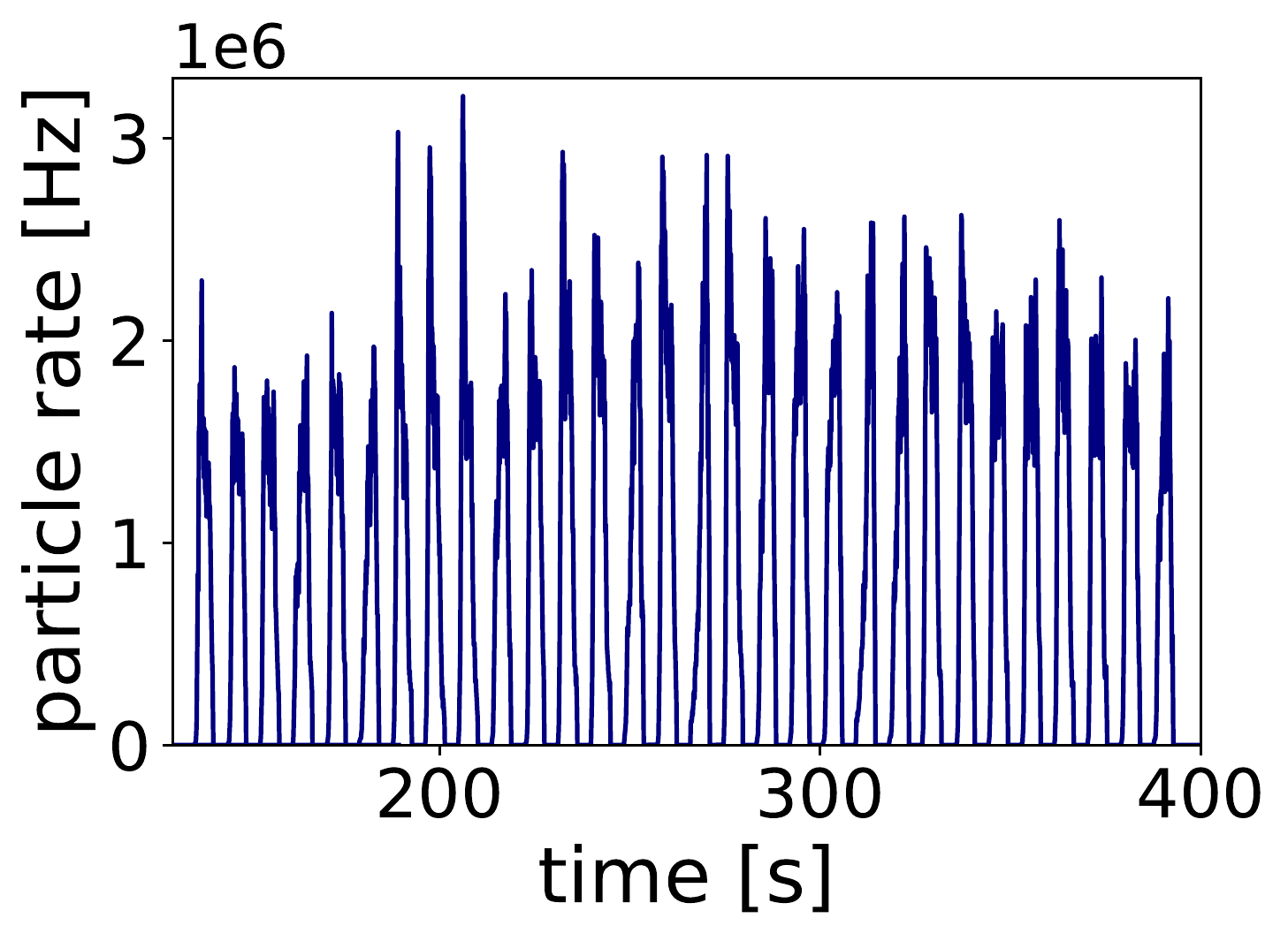}
		\caption{``high (p800)'' setting}
		\end{subfigure}
		\caption{Comparison of the spill structure of low flux protons for $\SI{800}{MeV}$ protons.  Setting ``low (p800)'' shows larger fluctuations in measured particles per spill compared to setting ``high (p800)''.}
		\label{fig:800structure}
		\end{figure}

    The particle rates for each setting were measured with the RaMon system for seven different beam energies below $\SI{252.7}{}
    $ $\SI{}{MeV}$ and three different settings for $\SI{800}{MeV}$. The combined results are shown in Figure \ref{fig:rateresults}. 
    \\
	The mean rate for the lowest medical flux setting, called ``low (MED)'', ranges from $\SI{2.43}{}-\SI{3.35}{kHz}$. Depending on the particle energy, a mean rate between $\SI{232}{}-\SI{435}{kHz}$ was obtained for setting ``medium (MED)''. For setting ``high (MED)'', mean particle rates between $\SI{3.73}{}-\SI{5.21}{MHz}$ were measured. \\ 
	The measured rates for the $\SI{800}{MeV}$ settings showed similar results compared to the medical settings. A mean particle rate of $\SI{1.95}{kHz}$ for setting ``low (p800)'', $\SI{198}{kHz}$ for setting ``medium (p800)'' and $\SI{1.25}{MHz}$ for setting ``high (p800)'' were obtained. Setting ``low (p800)'' showed more significant rate fluctuations per spill compared to the other $\SI{800}{MeV}$ low flux settings. Figure \ref{fig:800structure} depicts the spill structure for setting ``low (p800)'' in comparison to setting ``high (p800)''. 
		\\
	\subsection{Spot size measurement}
	\label{sec:spotsizemeas}
\begin{figure}[h]
	\centering
	\includegraphics[width=0.45\textwidth]{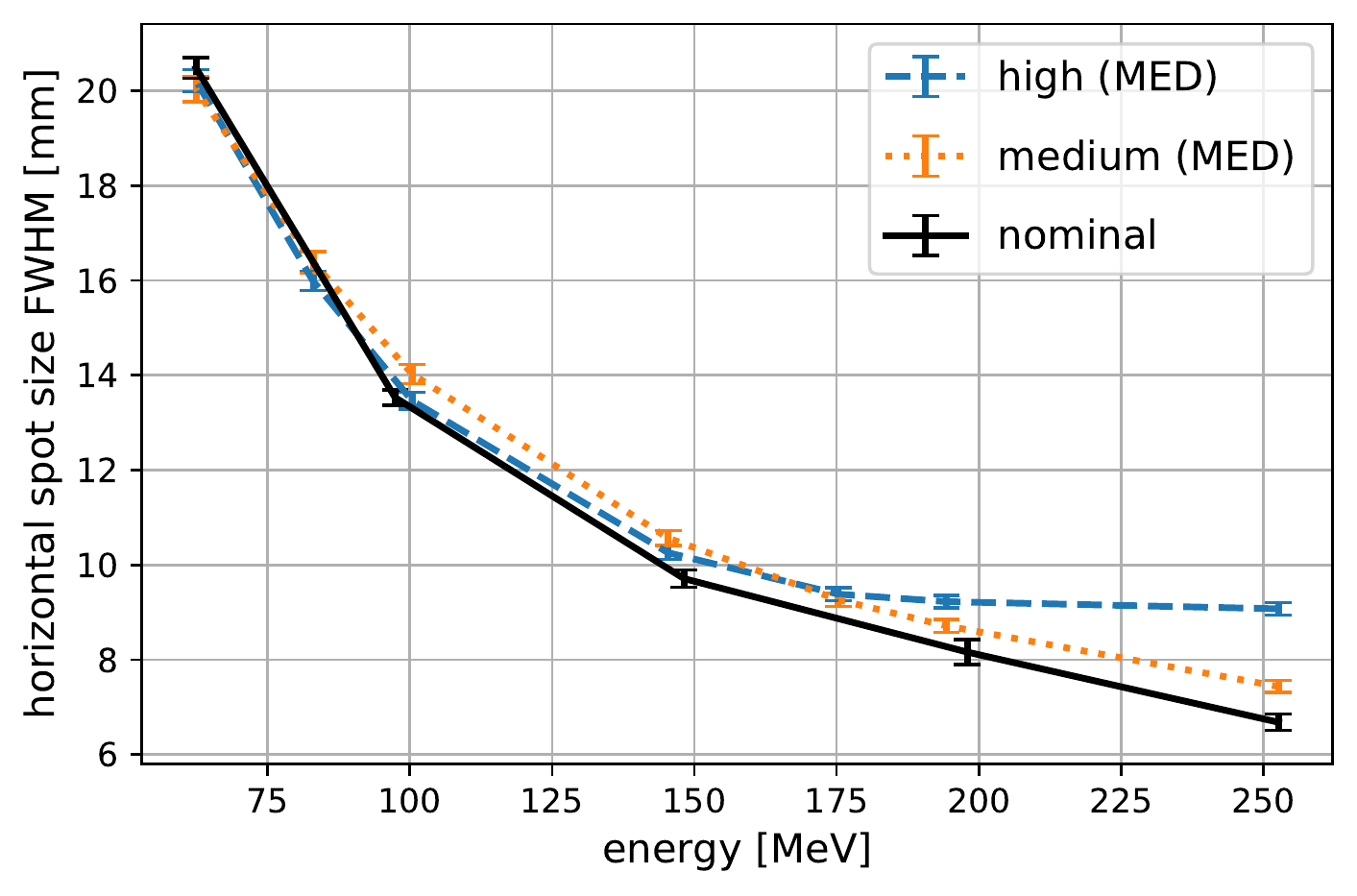}
	\caption{Horizontal (x-axis) beam spot size for different proton energies and particle fluxes.}
	\label{fig:spotx}
	\end{figure}

\begin{figure}[h]
	\centering
	\includegraphics[width=0.45\textwidth]{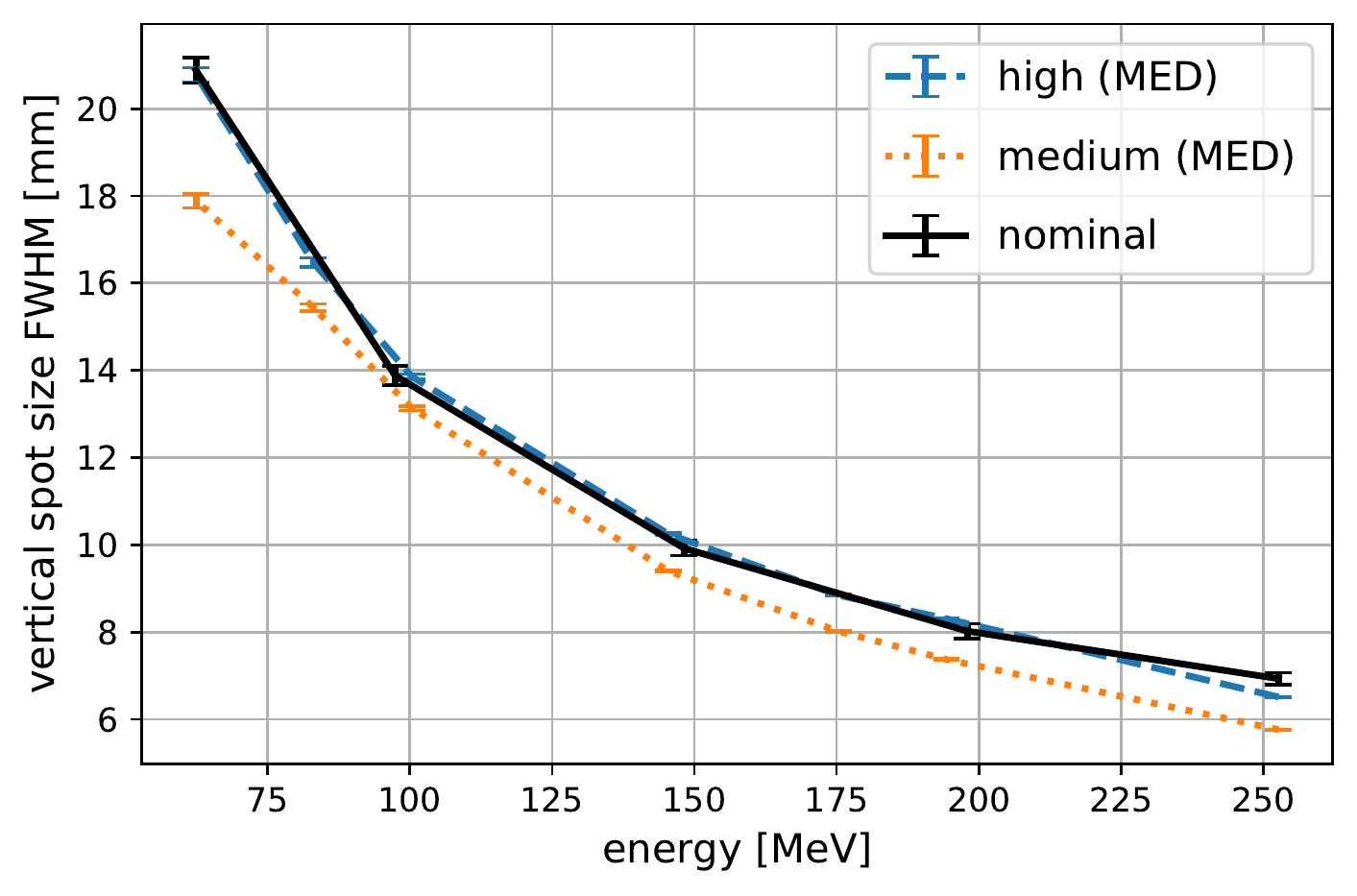}
	\caption{Vertical (y-axis) beam spot size for different proton energies and particle fluxes.}
	\label{fig:spoty}
	\end{figure}

	The Lynx\textsuperscript{\textregistered} detector was used to measure the spot sizes for the nominal setting ($\approx\SI{200}{MHz}$) in a prior measurement \citep{ir1abnahme}. A 2D Gaussian was fitted to the obtained 2D intensity profile distribution. The resulting full widths at half maximum (FWHM)  were compared to spot sizes of the low flux medical settings, which were measured with the DSSD. The beam spots for the ``high (MED)'' and ``medium (MED)'' settings also showed a Gaussian-shaped distribution. The measured horizontal spot size is depicted in Figure \ref{fig:spotx} and the vertical spot sizes in Figure \ref{fig:spoty}. 
	\\
	Similar vertical spot sizes compared to the nominal flux setting were obtained for the ``high (MED)'' setting. Except for the horizontal spot size for ``high (MED)'' at $\SI{252.7}{MeV}$, the absolute difference in FWHM compared to the nominal flux was always less than $\approx \SI{1}{mm}$ FWHM.
	The vertical spot size of the ``medium (MED)'' setting was slightly smaller compared to the nominal flux ($\approx \SI{1}{}-\SI{2}{mm} \text{ } \Delta$FWHM). On the other hand, the absolute difference in FWHM of the ``medium (MED)'' setting compared to the nominal flux did not exceed $\approx \SI{0.5}{mm}$ FWHM.\\
	
\begin{figure}[!h]
		\centering
		\includegraphics[width=0.8\linewidth]{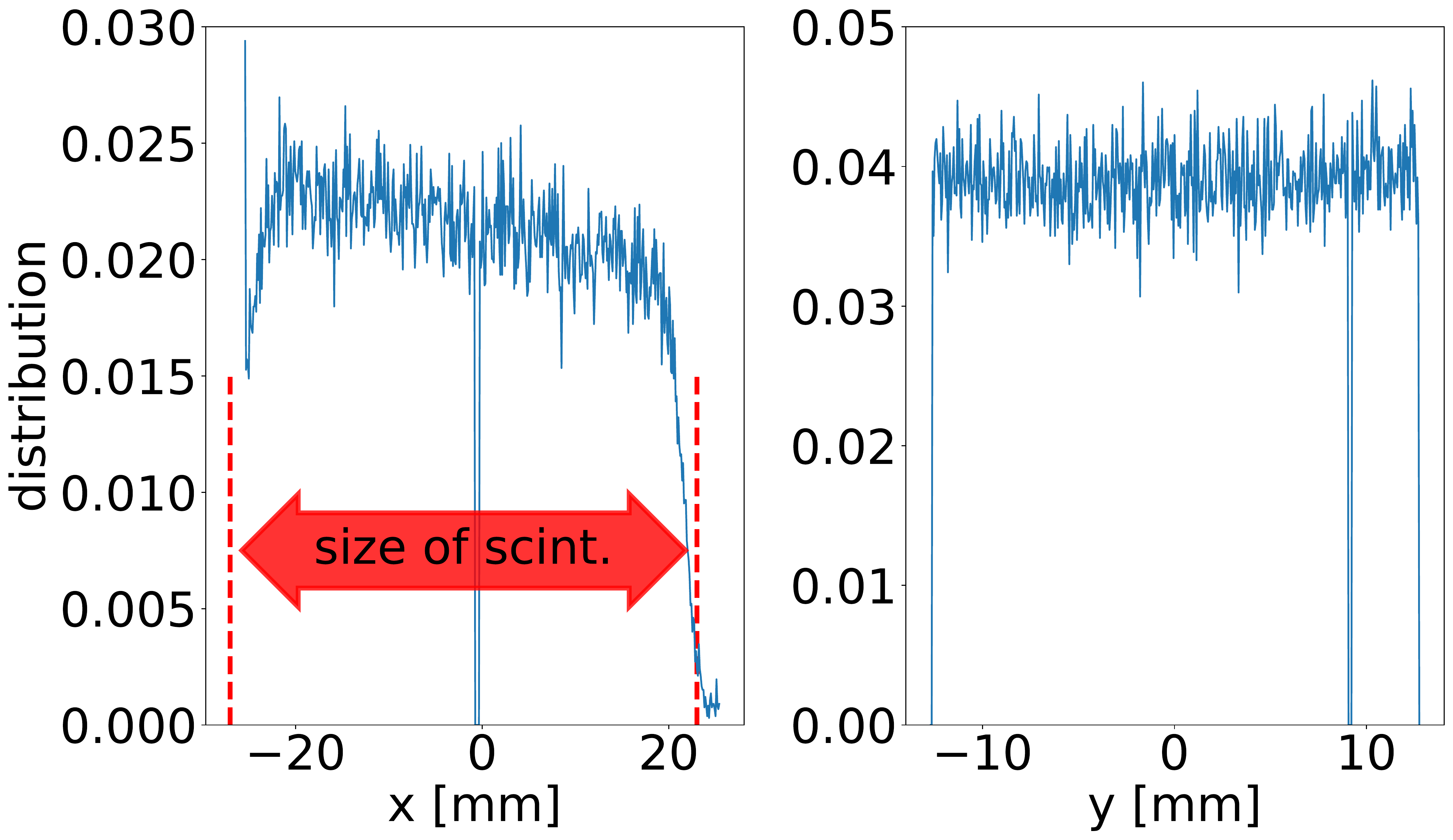}
		
		\caption{Horizontal (left) and vertical (right) beam profile of $\SI{252.7}{MeV}$ protons for setting ``low (MED)'' measured with a DSSD. The spot size cannot be determined since the beam is bigger than the trigger scintillators used for the DSSD. The beam profile of the horizontal beam close to the trigger edges is truncated due to the size of the trigger scintillators.}
		\label{fig:dssdresults}
		\end{figure}

\begin{figure}[!h]
		\centering
		\includegraphics[width=.99\linewidth]{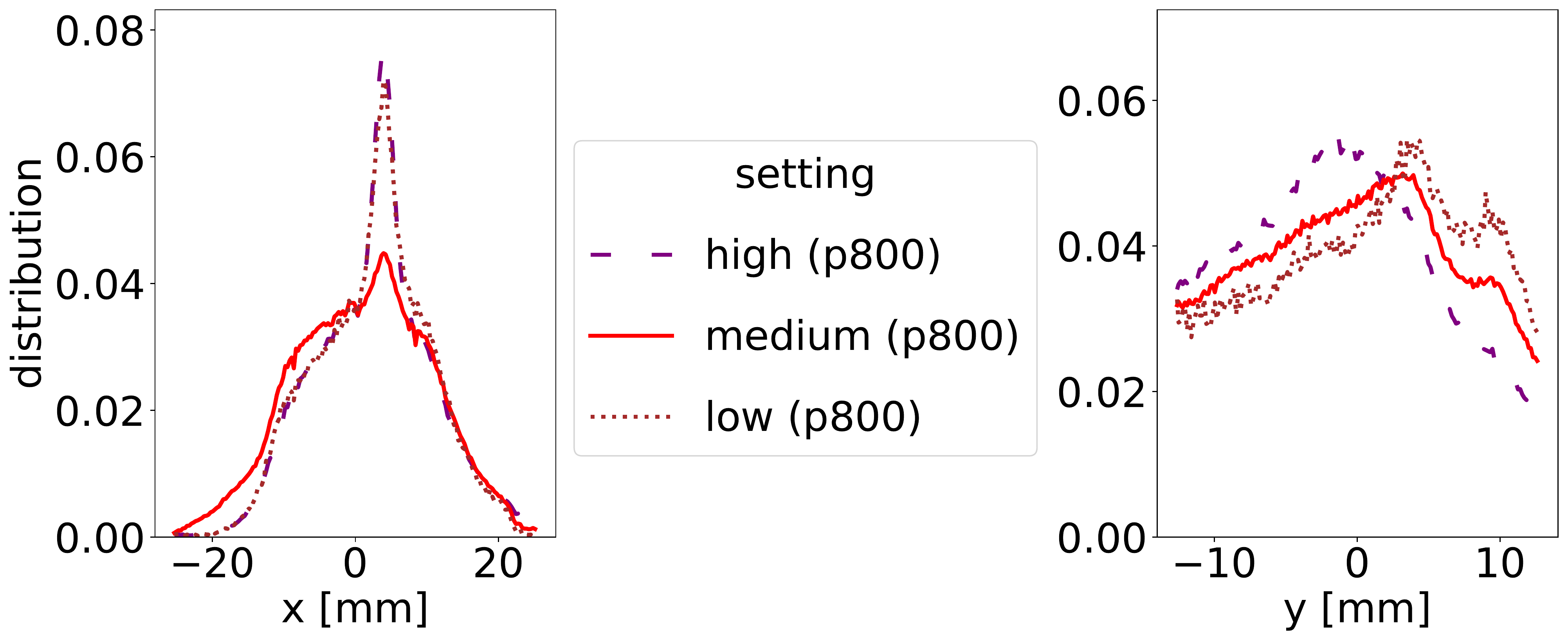}
		\caption{Horizontal (left) and vertical (right) beam profile of $\SI{800}{MeV}$ protons for setting ``low (p800)'', ``medium (p800)'' and ``high (p800)'' measured with a DSSD. The beam profiles do not follow a Gaussian distribution. The vertical beam is slightly larger than the vertical side of the DSSD.}
		\label{fig:800spots}
		\end{figure}

	The beam profiles of the ``low (MED)'' setting were also measured with the DSSD. Figure \ref{fig:dssdresults} shows the measured vertical beam profile of the ``low (MED)'' setting for $\SI{252.7}{MeV}$ protons as an example.	
	\\
		It is apparent that the shape of the beam is always bigger than the $\SI{5}{} \times \SI{5}{{cm}^2}$ plastic scintillators. Only events passing both scintillators are able to trigger an event in the DSSD, which leads to a truncated beam profile at the edges of the plastic scintillators ($\approx \pm \SI{25}{mm}$). This also means that the measured rate for this setting can only be related to the area of the plastic scintillators, which has to be taken into account when using this setting. The broadening of the beam is a direct result of the use of the magnetic quadrupoles in the HEBT. Due to the geometry of the trigger scintillators and the DSSD, only part of the beam profile distribution could be measured. However, the beam distribution appears to be relatively uniform inside the measured area ($\SI{5}{} \times \SI{2.56}{{cm}^2}$). This was also observed for the horizontal beam profile.
        \\\\
        The spot size for $\SI{800}{MeV}$ protons was also measured with the DSSD. Prior to the spot size measurement, the horizontal and vertical offset of the $\SI{800}{MeV}$ beam was corrected as described in Section \ref{sec:steering}. In contrast to the medical settings, the spot sizes for this energy were not measured in the isocentre, since they were recorded simultaneously to the rate measurement with the RaMon in the isocentre. For this purpose, the DSSD was placed $\SI{16}{cm}$ downstream the RaMon. Thus, the measured spot size is increased due to multiple Coulomb scattering in the plastic scintillators of the RaMon setup in front of the DSSD. The beam profiles for all three $\SI{800}{MeV}$ low flux proton settings are depicted in Figure \ref{fig:800spots}.
        \\
		From Figure \ref{fig:800spots}, it is apparent that the beam profiles for all $\SI{800}{MeV}$ settings do not follow a Gaussian distribution. A second peak near the centre of the spot could be detected for all settings. The vertical beam profile for all settings is slightly larger than the p-side (vertical axis) of the DSSD. On the other hand, the horizontal beam profiles are smaller than the n-side (horizontal axis) of the DSSD and appear centred.
		In order to estimate the spot size, the sample standard deviation ${\sigma}_{\mathrm{x}}$ of the horizontal beam profile was calculated. It resulted in $\SI{8.2}{mm}$ for the ``high (p800)'', $\SI{9.4}{mm}$ for the ``medium (p800)'' and $\SI{8}{mm}$ for the ``low (p800)''  setting. 
		
    \subsection{Energy measurement}
\begin{figure}[h]
	\centering
	\includegraphics[width=0.5\textwidth]{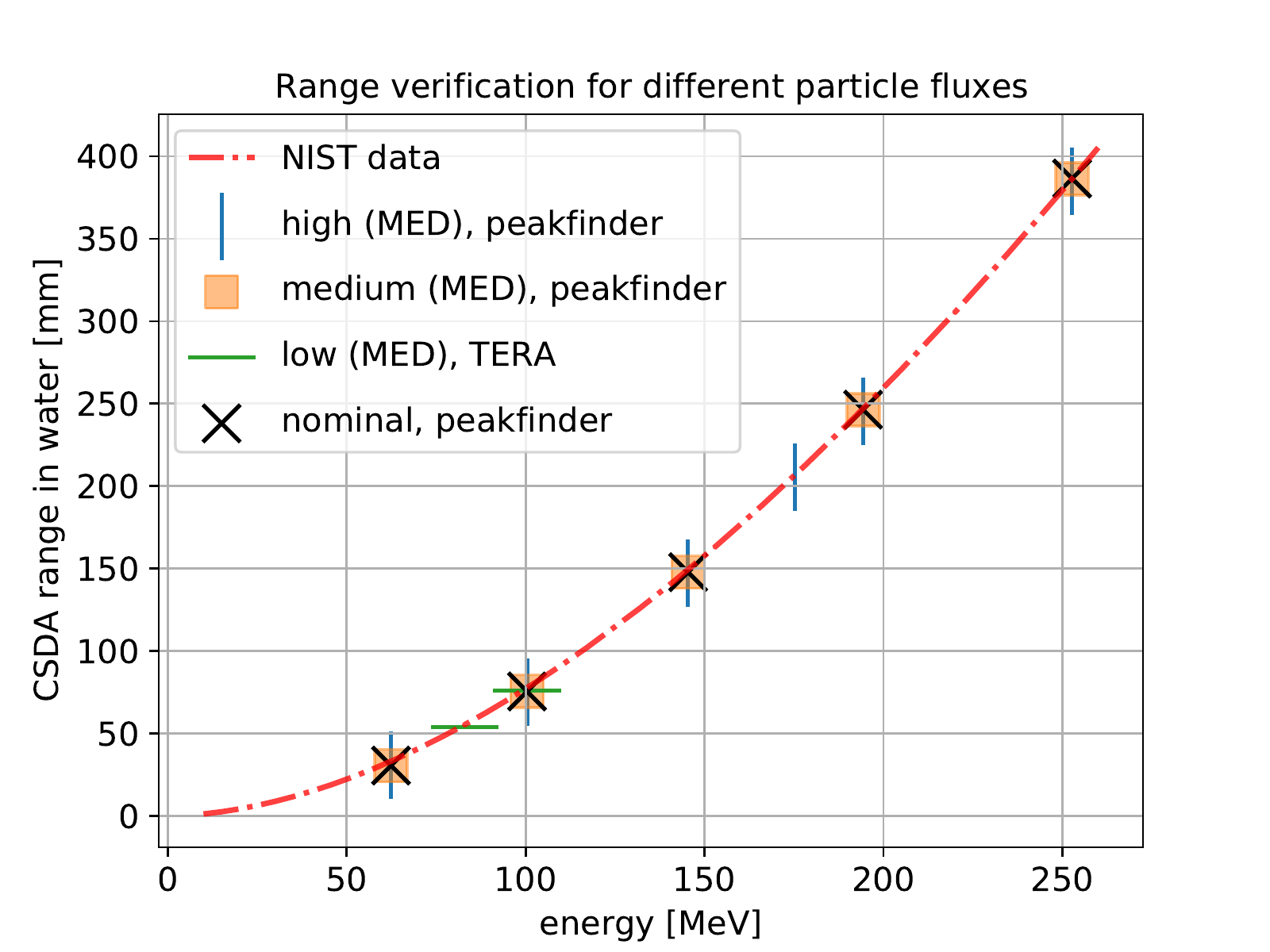}
	\caption{Measured range in water for different proton energies and flux settings. The obtained ranges were compared to data from the NIST PSTAR database.}
	\label{fig:rangeresults}
	\end{figure}

    The effect of the particle flux reduction method on the mean energy was studied by measuring the range in water for different flux settings and different beam energies. The measured range in water was then compared to the NIST PSTAR database. Because of an insufficiently low signal-to-noise ratio in the ionization chambers of the PTW PEAKFINDER\texttrademark{}, the TERA range telescope was used to measure the range in water for the ``low (MED)'' setting at $\SI{83}{}$ and $\SI{100.4}{MeV}$. For all other flux settings, including the nominal flux, the PTW PEAKFINDER\texttrademark{} could be used. As can be seen in Figure \ref{fig:rangeresults}, no significant change in the measured range was observed for any flux setting. The measured ranges are also in good agreement with the ranges obtained from the NIST database. \\In contrast to the rate and spot size measurements, the energy measurements were recorded prior to a timing optimization of the accelerator cycle to reduce dead time. Even though no change in the particle energy was expected, the measurement for the ``medium (MED)'' setting at $\SI{252.7}{MeV}$ was repeated after the timing optimization. As expected, the same range was obtained.

    \section{Summary and Outlook}
  In total, six low flux settings for protons were commissioned at MedAustron and can now be used in the non-clinical research room. Because the particle rate of the commissioned settings was well below the design specifications of the accelerator and beam diagnostic elements, the beam for the required settings had to be set up ``blindly'' without the standard beam instrumentation, only using external detectors. The beam energy, particle rate and spot size were measured for three settings at seven different beam energies ranging from $\SI{62.4}{}$ to $\SI{252.7}{MeV}$. Furthermore, three low flux settings were commissioned for $\SI{800}{MeV}$.  \\ A dedicated rate monitor system (RaMon), consisting of two plastic scintillators, was developed (Figure \ref{fig:ds}) for the particle rate measurements. For each energy, a stable setting with a mean particle rate of $ \mathcal{O} \left(\SI{}{kHz} \right)$, another with $ \mathcal{O} \left(\SI{100}{kHz} \right)$ and a setting with $ \mathcal{O} \left(\SI{}{MHz} \right)$ was commissioned.\\
The beam profiles were measured using the Lynx\textsuperscript{\textregistered} for higher particle rates in the clinical energy range and a DSSD for all low flux settings. Setting ``high (MED)'' and ``medium (MED)'' showed similar beam profiles compared to the nominal ($\approx\SI{200}{}$ $\SI{}{MHz}$) flux setting. However, a slightly smaller vertical spot size was obtained for the setting ``medium (MED)'' ($\approx \SI{1}{}-\SI{2}{mm} \text{ } \Delta$FWHM)). A  difference in the width of the measured horizontal beam profiles larger than $\SI{1}{mm}$ FWHM was only observed for the ``high (MED)'' setting at $\SI{252.7}{MeV}$. The effect of the magnetic quadrupoles used for the ``low (MED)'' setting resulted in a large beam spread, hence the full beam profile could not be measured for this setting. The measured rate for setting ``low (MED)'' can only be defined on the area as large as the plastic scintillators ($\SI{50x50}{} \SI{}{{mm}^2}$). The initially measured beam profiles for all $\SI{800}{MeV}$ settings showed a vertical and horizontal offset in the isocentre. Therefore, a beam position measurement based alignment had to be performed prior to the spot size and rate measurements for all p800 settings. This was done solely with external detectors (DSSDs) since the standard beam diagnostic elements in the HEBT were not designed for particle rates well below the therapeutic intensities ($\approx \SI{E3}{}-\SI{E6}{particles/s}$).
For all three $\SI{800}{MeV}$ settings, the measured beam profiles did not follow a Gaussian distribution. To estimate the spot size, the sample standard deviation was calculated for the horizontal beam profile. Depending on the setting, a sample standard deviation ranging from $\SI{8}{}-\SI{9.4}{mm}$ was obtained. In order to measure the full vertical beam profile for $\SI{800}{MeV}$, bigger detectors, able to handle low particle fluxes, are required. 
    \\ The range in water for different beam energies and flux settings was measured using the PTW PEAKFINDER\texttrademark{} and a range telescope based on plastic scintillator slices coupled to SiPMs. A comparison of the measured ranges in water for different particle energies and particle fluxes yielded that the particle flux reduction methods had no significant impact on the particle energy.
    \section*{Acknowledgements} 
    This project has received funding from the European Union’s Horizon 2020 research and innovation programme under the Marie Sk{\l}odowska-Curie grant agreement No 675265 and from the Austrian Research Promotion Agency (FFG), grant number 869878. 

    \bibliography{PaperNonClinical}

\end{document}